# Political problems facing governments in using scientific advice, and legal and other problems facing scientists when trying to give independent advice to governments. How effective are the Amaldi and Pugwash Conferences?

Klaus Gottstein[*], Nele Matz[**]

The rapid development of the modern world in regard to *inter alia* technical and technological capabilities results in a particularly high demand for well-informed political decision-making. This should be based on the setting of legal rules, regulations and standards. Modern nations and their governments as well as regional groups and organisations and the world community as a whole are faced today with severe security problems caused by the threats of nuclear proliferation and of the misuse of nuclear materials, by wars and civil wars and the weapons trade, by terrorism, nationalism, racism and ethnic "cleansing", the migration of millions of people, food and energy shortages, psychological, social, and economic instabilities, and environmental pollution, to mention only some of the most pressing ones. The solution of these problems requires not only diplomatic skill, sound judgement and good will but also a thorough knowledge of certain scientific facts, correlations and interconnections. Decisions and regulations in fields such as nuclear, biological and chemical weapons, civil use of nuclear power, climate change mitigation, biotechnologies, biosafety and biomedicine all depend upon expert advice in order to reach reasonable and well-balanced solutions. The scientific community which administers and continuously reviews and enlarges the body of knowledge should, and usually does, feel responsible for supplying the advice needed by political decision-makers. While the balancing of potentially diverging positions and interests in decisions is the primary function of the political decision-maker, it is the role of the scientific community to provide for the relevant knowledge on facts and capabilities and thus establish a sound background for the decision-making process. The complexity of the issues at stake requires a process of permanent research and expert advice to governments and international institutions to keep pace with developments, and to rapidly react to changes. To enhance legitimacy, expert advice must be organised in a way that safeguards its independence from vested interests.

In Germany, the German Research Society (Deutsche Forschungsgemeinschaft) sees as its task not only the promotion and support of science but also the supply of advice to government, parliament and the public in scientific questions. The Union of German Academies of Sciences, representing seven regional academies, has just proposed to set up a "Convent of German Science and Letters" as an advisory instrument in questions relevant to society. Governments and individual government ministries in many countries, including Germany, as well as the United Nations and its suborganisations, have appointed science advisers and set up scientific advisory committees of various kinds which study the issues at stake and develop options for action.

---

[*] Max Planck Institute for Physics, Munich
[**] Max Planck Institute for Comparative Public Law and International Law, Heidelberg

This does not mean that the best advice is always available everywhere and is heeded by those concerned. If it were, the state of affairs on this planet would be different, although the complexity of many problems and the inherited traits of human nature will always set limits to what the "best advice", whatever its definition, can do. Nevertheless, there is widespread criticism of the way in which government-appointed advisory bodies operate. Certainly, they are often helpful in finding answers to clear-cut questions, but they are usually content with just this task. In many cases, they are not authorised to supply answers to questions they have not been asked. On the other hand, it is often necessary to take a long-range view and tackle questions which the politicians have not yet recognised or, because of their potential preoccupation with winning the next elections, do not consider useful for this purpose. For this reason some scientific academies and societies, and various associations have set up committees and have organised conferences on issues of public interest, in order to make available information on the state of affairs in particular fields, on existing risks and long-term consequences, of measures taken or planned, and on alternative options. Frequently international co-operation is sought for the development of solutions for global problems. The Pugwash Conferences and the Amaldi Conferences are of this type. Their preoccupation from their very beginnings has been with the threat of nuclear weapons. Later other weapons of mass destruction and other threats to global security were included in their agenda.

Unfortunately, it must be stated that much of the relevant knowledge that is obtainable is not used by those who are meant to use it, either justifiably so because the advice based on it starts from unrealistic premises, or for reasons of negligence and poor management: because the advice does not reach the decision-makers in time, is not presented in a way conducive to application, or is incomplete in one way or another so that its use would lead to undesirable long-term results.

In this paper we shall try to consider the various reasons which make the production and the implementation of valid scientific advice to political decision-makers and to the public such a difficult undertaking.

1. Who is able to give scientific advice?

National Academies and National Scientific Societies usually have among their members the best experts in any field. However, the problems often require an interdisciplinary approach, and the experts are not necessarily able or willing to look beyond the borders of their field of specialisation. Interdisciplinary co-operation is sometimes not easy to organise. The guiding principle should be, of course, to include all the disciplines of knowledge whose expertise is needed for assessing the existing situation and the options available for action with regard to their respective risks and benefits. One difficulty here is that different disciplines often have developed their own special ways of thinking and of expressing themselves so that the same terms may have different meanings in different fields of scholarship. Sometimes, however, the need is felt to check, or supplement, the work of government-employed experts by an independent group of scientists who may belong to the same disciplines as the former but may start from different assumptions.

Sometimes a private group of scientists takes the initiative when its members see the need for it. Examples are the Pugwash and the Amaldi Conferences, the Union of Concerned Scientists, the Federation of American Scientists, the Federation of German Scientists (VDW) and many others. In some cases membership is by invitation only, in others anybody can join who is associated with a scientific institute. Not all of them have the same standing in the public perception. Statements by Nobel Prize Winners find particular attention by the media, even if the laureates express views on issues outside their own field of expertise.

This shows the need to distinguish carefully between private views expressed by ordinary citizens, even when they are a group of particular professional distinction, and carefully researched statements by experts in the fields relevant for the problem under consideration.

A special problem is the availability of independent experts in technically very complex areas where virtually all experts are government-employed. Cases of this type arise, for instance, in the fields of nuclear weapons and of space technology. Here it is the responsibility of academies and scientific societies to create committees of independent scientists with a professional background enabling them to study, understand and, if necessary, criticise the work done by the government experts.

Occasionally, the problem of co-operation between different scientific organisations arises. There may be suggestions for joint action in order to maximise public attention for their critical stance. On the other hand, in order to preserve their independent reputation, well-known organisations are often reluctant to join campaigns not initiated by themselves.

2. Why are governments sometimes reluctant, or not ready, to accept well-founded advice?

The reluctance of governments to listen seriously to experts is widespread. Most politicians prefer to follow their own plans and to listen to scientific advice only if it agrees with these plans of their own. Advice is unwelcome if accepting it means abandoning a course originally chosen. Why is this so? It is partly the fault of the governments and partly the fault of the experts. The political advice offered by non-political experts is often ignored by politicians for the following reasons:

a. Accepting the advice might require unpopular measures which, in turn, might jeopardise the outcome of the next election.

b. The advice given might have been espoused and advocated by the opposition party. Accepting it might look like yielding to the opposition.

c. The explanation offered for the advice may seem unintelligible, or there may be no plausible explanation at all accompanying the advice. In this case the suspicion may be justified that the advice given is based on doubtful assumptions or is aimed, in a one-dimensional manner, at just one single goal, without assessing properly the by- and after-effects that cannot be excluded. But often it is just these by- and after-effects which turn out to be of paramount political importance.

d. The advice given may not have taken into account all aspects which successful politicians have to keep in mind, so that the latter felt entitled not to take serious what they were told by the experts of individual disciplines. There may not have been an all-encompassing, interdisciplinary body which co-ordinated the "sectoral" advice given by the experts of the various isolated disciplines.[1]

---

[1] To avoid a deficit of this kind, the American Association for the Advancement of Science (AAAS) has recently established a Center for Science, Technology nnd Security Policy. This Center is intended to serve as a point of contact for Congress and the executive branch on the one hand and university-based antiterrorism research centers on the other

e. The advice may have been delivered at an inopportune time, or to the wrong address. It is a well-known problem for scientists to obtain enough knowledge of details of the political machinery to know where and by whom at which level the relevant decisions are really prepared, what is at the top of the current political agenda in relation to the question to be tackled, what are the perceived political or economic obstacles to the solutions proposed, and which options exist for overcoming the obstacles.

As a general rule scientific advisers to governments should refrain from suggesting what should be done unless they are told precisely which goal is to be reached and which boundary conditions are to be observed in approaching this goal. What should be done is a political, not a scientific decision. It often occurs, however, that scientific institutions, instead of offering a survey of the different options available for coping with a given situation, instead of estimating conservatively the risks, benefits and costs for each option, including the potential long-term consequences of side- and after-effects in other areas, limit themselves to isolated warnings and to recommendations as to what should, or should not, be done. In the absence of a complete picture in the light of all disciplines, it is then easy for politicians to evade the issue by concentrating on an area where the criticised measures have positive effects (for instance, by considering that disregard for environmental protection has short-term economic advantages).

With the assistance of science and technology it would no doubt be feasible in many cases to foresee the occurrence of by- and after-effects of human activities. They could be taken into account and made part of the overall planning. Countermeasures could be prepared, or the activities planned could be replaced by others with less harmful consequences. As the main effects and the by-effects often occur in different fields of specialisation (in security policy and in psychology, for instance, or in economy and in climate research), multidimensional thinking and interdisciplinary collaboration are required. Because of the global character of many problems, international co-operation is also indicated.

When scientists and scholars forecast what is going to happen they should clearly state under which assumptions they have come to that particular conclusion, and what the uncertainties are. In their advice to governments and the public they should limit themselves generally to explaining what the options for action are that could be followed, and what the potential risks, costs and benefits are that each option entails. These rules are often ignored so that decision-makers do not feel motivated to consider in earnest the advice given.

On the other hand, scientists and scholars should not wait until they are called upon to give advice. When signs of danger become visible in public developments which seem to go unnoticed by the authorities then it becomes the duty of the institutions of science to raise their voices, point out the facts and their potential consequences and specify the available remedies, if there are any.

3. At which levels is scientific advice needed?

Tasks for scientific advice can be found on four levels:

---

hand. Its role will be both informational and catalytic. It will respond quickly to requests for scientific information. For policy-makers needing scientific data on security issues related to antiterrorism the Center will offer "one-stop shopping". (PHYSICS TODAY, July 2004, page 30).

a. the level of science and letters: it may be necessary to have knowledge of the nature of certain problems, of latent and open conflicts, of their origins, and of the possibilities for dealing with them;

b. the level of practical policy: options for concrete action have to be worked out, with costs, benefits and risks assessed for each option, as mentioned in 1;

c. the level of the media: politicians and the public have to be informed about the situation so that the required political measures - even if unpopular at the outset - get the necessary support;

d. the level of education: the young generation has to be kept informed about the state of global affairs so that future leaders are prepared for the tasks they will have to tackle. Knowledge is not inherited, it must be taught and learned.

Scientists and scholars are able, and therefore have a special responsibility, to work at all of these four levels.

4. What are the channels through which scientific advice is made available?

Governments and the public can obtain scientific advice either from documents prepared by scientific institutions within the government and by government-appointed experts, or from memoranda and public statements issued by scientific institutions, by NGOs or by individual groups of scientists. Proceedings of scientific conferences on issues of public concern are also potential sources of information and advice useful for decision-makers. The readiness of governments to accept advice "from outside", i.e. from non-governmental sources, is often rather limited and may have to be enhanced by special lobbying.

It is the task of the media to inform the general public about the current issues and the available options for action.

It is an important question by which legal means the independence of scientific advisers to governments, official advisers as well as non-governmental ones, can be secured in the interest of truthful, unbiased information on the potential risks associated with human activities and in particular with government policies. The discussion on legal questions of independent scientific advice to governments has to distinguish between different structural forms of expert advice, in particular between, firstly, institutionalised bodies of experts that are established by statute or decree to advise governments and, secondly, publicly and/or privately funded research institutions that function as "think-tanks", and, thirdly, experts who are asked for their opinion more or less informally by certain politicians, and, fourthly, concerned members of the scientific community who contribute to the public debate without request by members of the government.

5. The legal relationship between governments and scientific research

The legal relationship between the state and scientists is governed by a conglomerate of rights, freedoms and obligations. First of all, a distinction has to be drawn between substantial constitutional law that, at least in democratic states, safeguards the principle of scientific freedom on the one hand and procedural issues on the other, concerning, for example, how to establish an independent expert commission by law. The substantial freedom of research is a *conditio sine qua non* for independent expert advice to governments. However, the freedom of scientific research does not go completely unregulated or unlimited despite its constitutional guarantees. Laws and regulations might either prohibit scientific research in certain branches of science, or political decisions may regulate research by directing public funds. The "soft" regulation by the targeted direction of funds is the regular practice, while, due to constitutional restraints, the legal prohibition of research must be considered the exception. Such exceptions as in the case of the prohibition of reproductive cloning of humans and related regulations on research on human embryonic stemcells raise legal difficulties and public dispute.

6. Guiding principles to safeguard quality and independence of expert advice

The quality and independence of research and advice is influenced by a variety of factors. The central issue is to provide for a broad basis of expert opinion from different disciplines or branches of research, to establish principles for their inclusion in decision-making processes and to keep expert advice unbiased. Naturally, not all issues subject to decision-making can be supported by expert commissions, since the decision-making process would become intolerably slow. A line has to be drawn between issues that are as important as they are difficult to deal with by decision-makers. For priority issues in need of multidisciplinary approaches expert commissions should be established by law or decree explicitly safeguarding independence and transparency regarding procedures for appointment and length of the term of service on the board. Further procedural details can be regulated by the statutes of the relevant meeting. Another distinction has to be made between expert advice on the national level and expert advice to international institutions because the structural setting and legal framework governing the decision-making process is quite different in the two cases. However, some principles that shall govern expert advice can be generalised.

a. Transparency and accountability

The overriding principle of high-quality and independent expert advice is transparency. This principle which can be guaranteed by rules and regulations should apply to all aspects of the selection process, in particular who is asked for advice, by which procedures advice is obtained, what recommendations were given and how politics reacted to the recommendations. The principle of transparency enhances legitimacy of the decision reached by the government.

b. The multidisciplinary approach

As mentioned above, many issues of modern decision-making require input from more than one branch of science. A multidisciplinary approach is not restricted to issues where natural and social scientists shall work together but also refers to different branches of natural sciences. If, for example, advice on weapons of mass destruction is required it will in general not be sufficient to consult experts on nuclear physics. In general, it must be safeguarded that all relevant disciplines are represented in the relevant body that functions as a regular advisor or in ad-hoc expert hearings.

c. Nomination

If expert advisory commissions are established, the choice of their members is particularly important. Regulations can serve to prevent that experts are members of the legislative or have close industrial ties. It may be desirable, for the sake of independence and transparency, to invite proposals from the leading national research institutions, and then appoint members from the list of their suggestions.

7. Advisory commissions in international institutions

In principle the same considerations to keep expert advice unbiased and of high standard that are valid on the national level also apply to the international level. Most multilateral environmental agreements, e.g. the Convention on Biological Diversity (CBD) and the Framework Convention on Climate Change have established expert bodies for technical advice that report and give recommendations to the Conferences of the Parties. A lack of transparency and, potentially, a lack of quality and independence, however, results from the fact that the bodies consist of appointed government representatives with expertise in the relevant fields. The process of appointment by the national governments is not regulated. Furthermore, as government representatives the experts are not necessarily expected to discuss issues independent of their government's political stance.

8. What can the scientific community do to supply useful advice to decision-makers and the public in questions of general concern?

The first step, obviously, should be some sort of stocktaking. There should be a careful analysis of:

   a. the measures that are already under way to deal with the problem under consideration. (Are they appropriate and are they sufficient, are they all that can be done?),

   b. the results obtained so far,

   c. the difference between the results obtained and the results desired,

   d. the reasons for the difference. Are there reasons to believe that better results could have been obtained, had certain recommendations been taken into account? If so, what were the obstacles to implementing the recommendations?

The next step should be to consider carefully the quality of the existing guidelines or recommendations and the possibilities for improving them, assessing in advance the potential obstacles to their implementation as well as the existing ways and means for overcoming the obstacles. Questions to be asked in this context are the following:

   i. Are the scientific recommendations given, or to be given, presented in a form convincing to decision-makers?

   ii. Are they based on interdisciplinary research including economists, historians, psychologists, political scientists, etc.? Or do the recommendations, for instance, neglect economic, historical, psychological, political aspects or obstacles that should have been taken into account?

   iii. Are there gaps in knowledge which could be filled by more research?

   iv. Was there a lack of publicity so that the available options and their risks and benefits did not become known sufficiently well to the public and to the decision-makers?

v. Is there a lack of political will discernible because the available recommendations are unpopular due to insufficient explanation of the alternatives?

Usually, there exist several possible solutions, each with its specific advantages, disadvantages, risks and costs. Which of these solutions seems to be preferable in security policy, depends on the interests of the parties negotiating with each other. An obvious question is: Are the benefits, risks and costs associated with any one solution distributed equally among the negotiating parties and/or the governments and nations they represent, or will one party bear most of the burdens while another one will enjoy most of the benefits? In the name of intergenerational equity decision-makers who are aware of their long-range responsibilities should ask: Will only the present generation have the advantages from the chosen solution whereas future generations will have to pay the costs?

9. What is involved in evaluating different options for action?

The necessity to investigate different options for the solution of problems will require finding carefully balanced answers to the questions under consideration, a thorough analysis of the relevant facts and developments, their interconnections and trends, and of the probabilities and uncertainties involved. The preparation of detailed options may involve lengthy, interdisciplinary investigations. To choose among the options will be the task of the decision-makers in the centres of political power or of their deputies at the negotiation table. The scientists can only tell them what the options are, indicating the long-range risks, costs and benefits to be expected for each option. Given the complexity of many problems on the agenda of today's international negotiations, a considerable number of preparatory meetings of scientific experts is often required before clear-cut alternative options can be presented to the political decision-makers.

Whenever a technical solution that seems necessary from an overall and long-range point of view turns out to be unacceptable politically at the time being, it will be necessary to determine carefully the nature of the obstacles. Scientists sometimes do not have sufficient access to the realities of political life so that their advice disregards human and political factors and thus becomes unusable. This is why the analysis of the reasons for the non-implementation of scientifically convincing recommendations is so important. The next step should then be to work out soberly and unpolemically what the options are for overcoming the obstacles, and in which way the public could be informed and educated so that the removal of the obstacles becomes feasible politically. If necessary, the public must be convinced that the advice given on scientific grounds is sound, and that the politicians in charge should follow it if they do not want to lose public support.

10. How effective are the Amaldi and the Pugwash Conferences?

The purpose of the Pugwash Conferences is "to bring together, from around the world, influential scholars and public figures concerned with reducing the danger of armed conflict and seeking co-operative solutions for global problems. Meeting in private as individuals, rather than as representatives of governments and institutions, Pugwash participants exchange views and explore alternative approaches to arms control and tension reduction with a combination of candour, continuity, and flexibility seldom attained in official East-West and North-South discussions and negotiations. Yet, because of the stature of many of the Pugwash participants in their own countries (as, for example, science and arms-control advisers to governments, key figures in academies of science and universities, and former and future holders of high government office), insights from Pugwash discussions tend to penetrate quickly to the appropriate levels of official policy-making".[2] In the period of strained official relations and few unofficial channels during the Cold War, the fora and lines of communication provided by Pugwash played useful background roles in helping lay the groundwork for the Partial Test Ban Treaty of 1963, the Non-Proliferation Treaty of 1968, the Anti-Ballistic Missile Treaty of 1972, the Biological Weapons Convention of 1972, and the Chemical Weapons Convention of 1993. Subsequent trends of generally improving East-West relations and the emergence of a much wider array of unofficial channels of communication have somewhat reduced Pugwash's visibility while providing alternate pathways to similar ends, but Pugwash meetings have continued until the present to play an important role in bringing together key analysts and policy advisers for sustained, in-depth discussions of the crucial arms-control issues of the day. Since this is the aim of the Pugwash Conferences, it may be said that they are continuing to fulfil their purpose. But it is hard to say to what extent the results of the Pugwash meetings of the more recent years had an impact on the deliberations and decisions of decision-makers in governments.

Whereas the participants of Pugwash Conferences meet and discuss as private individuals, as mentioned above, not representing any organisation, the participants of the international Amaldi Conferences of Academies of Sciences and of National Scientific Societies on Problems of Global Security are delegates of the academies of sciences and national scientific societies of their home countries. The topics of these conferences are similar to those of the Pugwash conferences, they concern mostly questions of arms control and only to a small extent problems of security in a wider sense. The main purpose of the Amaldi Conferences is to keep the scientific community up to date with the developments in the production and control of weapons of mass destruction so that independent experts become available as counterparts to government experts in parliamentary hearings, public discussions, and in the media. Therefore, there is less emphasis on issuing public statements after Amaldi Conferences than there is after Pugwash Conferences. But the proceedings of Amaldi Conferences are made available to interested government departments and parliamentary committees, at least in Germany. In a few cases, foreign ministry officials and members of parliamentary committees have taken part in Amaldi Conferences. Nevertheless, there is certainly room for improvements in the effectiveness of the Amaldi Conferences, internally as well as externally: internally on the side of the academies and scientific societies in the intensity of their participation, and externally in the drive of their leaders for making available the expertise gained at Amaldi Conferences to those decision-makers on the national and European levels and at the United Nations who ought to use it. This would imply starting a learning process on how to overcome the obvious obstacles to a better communication between science and governance.

**DISCUSSION:**

KLAUS GOTTSTEIN – Just a brief reply to what Professor Haeckel said. Professor Haeckel, if I understood him correctly, asked: "Why do governments need scientific advice if they already know what they want to do?" That is usually the case, so why do they need advice?

---

[2] PUGWASH CONFERENCES ON SCIENCE AND WORLD AFFAIRS. (A Brief Description). Text issued by the Pugwash Central Office, London, September 1995.

Well, if governments know what they want to do, they often do not know how to achieve this purpose. If they have already a clear goal, like putting a man on the moon, they often still don't know how to do that. Then the scientists can help them to achieve that goal.

But it can also be that the scientists are forced to say: "This cannot be done." This was the case when President Reagan said he wanted to protect the United States from any enemy rocket hitting the area of the United States. The scientific answer was that some sort of anti-missile system can certainly be set up but it will only be possible to catch some of the incoming rockets, not all of them. To catch all of them would get so expensive that in practice it cannot be done.

So there are cases where the government knows what it wants to do and where scientists can tell them whether or not that is possible and what the consequences would be.

In the case of the reactor in Bavaria, near Munich, it was clear that the Bavarian government wanted the reactor, whereas the federal government, setting different priorities, was against it. But all scientists, on both sides, were agreed that such a reactor could be built. They only started from different premises, depending on the side which they advised. There are benefits and there are risks. One side claimed that the benefits outweigh the risks. The other side maintained that the risks, though small, were unacceptable. But that were political judgements, not scientific ones. That's why the scientists supporting the Bavarian government for the reactor and the scientists supporting the federal government against the reactor essentially agreed on the technical aspects, but they started from different assumptions about the acceptability of risks.

Then there are cases where the government does not know what to do, how to deal with a situation. An example is the present economic situation in the European Union and particularly in Germany.

As I said, scientific advisers should never say what a government should do, they only should show what the options are, with their benefits and their risks. In the economic case, for instance, there are several routes which the governments could follow. Each has benefits, chances and risks. The scientists, in this case economic scientists, technologists and so on, can just tell the governments for each of the routes what the possible consequences could be, what might happen with which probability (of course, nobody knows exactly what is going to happen, but you can estimate the probability). Then it might be that the probability of bad consequences (more unemployment, more deaths, for instance) is such that a government says: "We don't want to take that risk".

This is how scientific advice SHOULD be given. Unfortunately, scientists often make the mistake that they tell their government what it should do or what is going to happen without mentioning options and probabilities.

So I think Prof. Haeckel is wrong in assuming that no independent scientific advice is needed. It is needed in both of the following cases:
1.) The government knows what it wants, but it doesn't know how to achieve it.
2.) The government does not know how to deal with a certain situation. Then scientists might be able to show them the options, with the potential consequences (benefits and risks). Ideally, also with the consequences in areas other than those under present consideration, for instance in the social area, when a decision is taken in a field belonging to natural science.

As I mentioned in my talk, it may be that in order to protect the environment something has to be done that has short-term negative consequences for the economy, but in the long run is necessary because otherwise with the deterioration of the environment also the economy would suffer.